\documentclass[conference]{IEEEtran}
\IEEEoverridecommandlockouts
\usepackage{cite}
\usepackage{amsmath,amssymb,amsfonts}
\usepackage{algorithmic}
\usepackage{graphicx}
\usepackage{textcomp}
\usepackage{xcolor}
\def\BibTeX{{\rm B\kern-.05em{\sc i\kern-.025em b}\kern-.08em
   T\kern-.1667em\lower.7ex\hbox{E}\kern-.125emX}}
   
\usepackage{pgfplots}
\pgfplotsset{compat=newest}
\usetikzlibrary{plotmarks}
\usetikzlibrary{arrows.meta}
\usepgfplotslibrary{patchplots}
\usepackage{grffile}
\usepackage{amsmath}
\usepackage{pgfplots}
\usepackage{soul}

\usepackage{pifont}

\usepackage{multirow}
\usepackage{rotating}

\DeclareMathSymbol{\smi}{\mathbin}{AMSa}{"39}
\usepackage{textcomp}
\usepackage[absolute]{textpos}
\usepackage{amsmath,amsfonts,amssymb,amsthm, bm}
\usepackage{arydshln}
\usepackage{subfig}
\usepackage{adjustbox}
\usepackage{url}
\usepackage{atbegshi} 
\pgfplotsset{plot coordinates/math parser=false}
\newlength\figureheight
\newlength\figurewidth
\usepackage{caption}   
    
\newcommand{\at}{\makeatletter @\makeatother}

\AtBeginShipout{\AtBeginShipoutUpperLeft{%
		\put(\dimexpr\paperwidth-20cm\relax,-0.5cm){\parbox[t]{19cm}{\scriptsize \copyright
				2024 IEEE.  Personal use of this material is permitted. Permission from
				IEEE must be obtained for all other uses, in any current or future
				media, including reprinting/republishing this material for advertising
				or promotional purposes, creating new collective works, for resale or
				redistribution to servers or lists, or reuse of any copyrighted
				component of this work in other works.}}%
}}

\begin{document}

\title{Design Space Exploration at Frame-Level for Joint Decoding Energy and Quality Optimization in VVC\\
}

\author{\IEEEauthorblockN{Teresa St\"urzenhof\"acker, Matthias Kr\"anzler, Christian Herglotz, Andr\'e Kaup}
\IEEEauthorblockA{Chair of Multimedia Communications and Signal Processing, \\
Friedrich-Alexander-Universit\"at Erlangen-N\"urnberg (FAU),
Erlangen, Germany\\
\{teresa.stuerzenhofaecker, matthias.kraenzler, christian.herglotz, andre.kaup\}\at fau.de}
}

\maketitle

\begin{abstract}
% Introduction to topic - provide background
% Purpose, Gap in previous research
 In the pursuit of a reduced energy demand of VVC decoders, it was found that the coding tool configuration has a substantial influence on the bit rate efficiency and the decoding energy demand. 
The Advanced Design Space Exploration algorithm as proposed in the literature, can derive coding tool configurations that provide optimal trade-offs between rate and energy efficiency.
 Yet, some trade-off points in the design space cannot be reached with the state-of-the-art methodology, which defines coding tools for an entire bitstream.
% Novel idea of this work
This work proposes a novel, granular adjustment of the coding tool usage in VVC.
Consequently, the optimization algorithm is adjusted to explore coding tool configurations that operate on frame-level. 
Moreover, new optimization criteria are introduced to focus the search on specific bit rates.
 % Results, Conclusion
 As a result, coding tool configurations are obtained which yield so far inaccessible trade-offs between bit rate efficiency and decoding energy demand for VVC-coded sequences. 
The proposed methodology extends the design space and enhances the continuity of the Pareto front. 
%Additionally, the Pareto curve exhibits a downward shift by around 5\% $\text{BDDE}_{\text{VMAF}}$ in the region between 12-22\% $\text{BDR}_{\text{VMAF}}$.

\end{abstract}
\begin{IEEEkeywords}
VVC, Energy Efficiency, Optimization, Decoder
\end{IEEEkeywords}
\vspace{-0.3cm}
\section{Introduction}
%A significant increase in Internet Protocol (IP) traffic can be observed for video content over the past few years, and according to Cisco's Visual Networking Index, video content accounts for more than 80\% of the transmitted data in 2022.
%According to Cisco's Visual Networking Index a significant increase in Internet Protocol (IP) traffic could be observed for video content over the past few years \cite{CISCO}. 

A significant increase in Internet Protocol (IP) traffic could be observed for video content over the past few years.
The most recent statistics, gathered in Ericsson's mobility report from November 2023, demonstrate that in the year 2023, the transmitted video content already accounted for more than 70\% of the total mobile traffic \cite{ericsson}.
%The video content was projected to account for more than 80\% of the transmitted data in 2022 \cite{CISCO}.
%Todo Check reference again! - Ericson Report
This striking dominance is associated with the popularity of video conferencing tools and video on-demand services. Especially, the emerging trend for high-resolution content increases the data sizes drastically.
%Todo: potentielle quelle: \cite{impactvideoCarbon}
%The Versatile Video Coding (VVC) standard from 2020 provides an efficient way to compress this sheer amount of video data. With VVC, it is possible to achieve a bit rate reduction of 50\% compared to its predecessor High Efficiency Video Coding (HEVC), while preserving an equal subjective quality \cite{DevelopmentCoders}.
With Versatile Video Coding (VVC), an efficient way to compress video content was provided in 2020. Thereby it is possible to achieve a bit rate reduction of 50\% compared to its predecessor High Efficiency Video Coding (HEVC), while preserving an equal subjective quality \cite{DevelopmentCoders}.
The enhanced compression is largely attributed to new and advanced coding tools. However, these new compression techniques introduce additional computational complexity \cite{complexityVVC}. As a consequence of the increased complexity, more energy is required to perform encoding and decoding. 
Previous studies showed that the energy demand of VVC decoders is up to 80\% higher than that of comparable HEVC decoders \cite{compareEnergyandTime}, \cite{dse}, \cite{reviewSoftwareHardware}.
For video-on-demand applications, the energy demand at the encoder is negligible since a video is only encoded once and then distributed to servers or cloud-based storage systems.
%\cite{impactvideoCarbon}
In contrast, the decoding process is executed many times and by each end device \cite{TradeOffs_videoCodecs}. Thus, a low decoding energy demand is not only valuable to keep the overall energy consumption of the streaming process modest, but it also extends the battery life of mobile end devices with limited power supply.
The endeavour of a more frugal and economic management of resources in video streaming leads to the ultimate objective to find optimum compromises between decoding energy and compression.
%Therefore, the ultimate objective to find optimum compromises between decoding energy and compression can be derived from the endeavour of a more frugal and economic management of resources in video streaming.
In previous work, it was found that the coding tool configurations can be used to control this efficiency trade-off \cite{dse}. A greedy-strategy-based advanced design space exploration (ADSE) algorithm was used to solve this multi-objective optimization problem and identify coding tool profiles (CTPs) with joint energy and rate efficiency \cite{adse}. It was shown that encoding a sequence with such an optimal CTP reduces the energy demand at the decoder compared to the \textit{randomaccess slower} CTP of VVC.

%So far, two dominant efficiency trade-off clusters have been identified while investigating possible CTPs \cite{adse}. 
%The first cluster is located at around 10\% additional rate while saving 40\% of decoding energy a second cluster is found at 23\% additional bit rate and 50\% savings in decoding energy.
%In \cite{adse} it was identified that there are two dominant clusters in the design space.
In \cite{adse}, it becomes apparent that there are sparse regions, where no coding tool configuration can provide an efficiency trade-off. 
This can be reasoned with the scope of the definition of coding tool usage. 
In the VVC state-of-the-art coder, a coding tool can either be enabled or disabled for the entire bitstream.
To overcome this limitation, this work offers the following contributions:

\begin{itemize}
  \item Modification of VVC encoder to allow enabling/ disabling of coding tools on frame-level granularity.
  \item Adjustment of the DSE optimization algorithm to search the enlarged set of possible CTPs.
  \item Novel optimization criteria to focus the search of optimal CTPs on a specifiable region in the design space.
\end{itemize}

%Paper structure

In Section \ref{sec:ADSE}, the prevalent joint optimization of bit rate and energy efficiency with the advanced design space exploration (ADSE) is explained. The extension that allows for adjustment of the coding tool states on frame-level is described in Section \ref{sec:extension}.
This section includes the modifications of the VVC encoder and the extension of ADSE which is required to search for efficient CTPs with tool usage that is specified on frame-level. Moreover, alternative optimization criteria are proposed to specify the search in the design space, which is spanned by the decoding energy demand and the amount of compression.
In Section \ref{sec:setup}, the set-up and metrics are described. The evaluation of the results is presented in Section \ref{sec:results}. Finally, a short summary and an outlook to future research is given in Section \ref{sec:conclusion}.

\section{Optimizing Bit Rate and Energy Efficiency with the Advanced Design Space Exploration}
\label{sec:ADSE}

The energy that is needed to decode a VVC-coded sequence is closely related to the coding tool configuration at the encoder. Coding tool profiles (CTPs) which result in minimum decoding energy while maximizing compression efficiency are considered as optimal. A successful approach to obtain CTPs with optimal joint efficiency trade-offs is the advanced design space exploration (ADSE) presented in \cite{adse}.
A CTP can be described by a binary vector $\boldsymbol{u}$, where each element represents the usage characteristic of a coding tool (CT).
As baseline CTP, we use the \textit{randomaccess slower} profile of VVC and denote it with $\boldsymbol{u_{1,0}}$.  Accordingly, the reference CTP corresponds to the baseline configuration in the first iteration.
The ADSE is an iterative algorithm that creates a set $\mathbf{U}_i$ of variations from the reference CTP in each iteration $i$.
 In each derived CTP $\boldsymbol{u}_{i,\nu}$, the usage of the $\nu$-th CT in the reference configuration is inverted. 
Thereby, it is possible to investigate the isolated impact of a single CT on the given reference configuration.
 All the derived CTPs are tested by encoding a set of test video sequences. Subsequently, the energy demand at the decoder, the resulting quality, as well as the compression efficiency are evaluated for each tested CTP and compared against the performance of the reference configuration.
 
 The reduction of decoding energy is quantified with the Bj\o ntegaard-Delta decoding energy ($\text{BDDE}_{\text{VMAF}}$) and the level of compression is compared with the Bj\o ntegaard-Delta Bit Rate ($\text{BDR}_{\text{VMAF}}$). A more detailed description of the metrics is given in Section \ref{sec:setup}.
 The joint efficiency of a CTP $\boldsymbol{u}$ is evaluated by the cost function,
 \begin{equation}
 	f(\boldsymbol{u}) = \text{BDR}_{\text{VMAF}}+\text{BDDE}_{\text{VMAF}},
 	\label{eq:classicCostfunction}
 \end{equation}
 which is therefore also the optimization criterion and the metric for comparison. In \cite{adse}, this is referred to as \textit{combined all} (CA).
 CTPs which reduce the decoding energy while only requiring a minimal amount of additional bits compared to the reference CTP are selected.
The selection process can be described as follows,
\begin{equation}
	\mathbf{S}_i = \{ \boldsymbol{u}_{i,\nu} \,| \, f(\boldsymbol{u}_{i,\nu}) < f(\boldsymbol{u}_{i,0}) \, \text{ for each } \boldsymbol{u}_{i,\nu} \, \text{in } \mathbf{U}_i \} ,
	\label{eq:select}	
\end{equation}  
 where $\mathbf{S}_i$ denotes the set of selected CTPs that were found to be an improvement compared to the reference CTP $\boldsymbol{u}_{i,0}$.
 From the set $\mathbf{S}_i$, all the CT usage characteristics that were found to offer an improvement are combined to define a new reference CTP $\boldsymbol{u}_{i+1,0}$ for the next iteration $i+1$. The algorithm terminates if the same reference CTP is obtained in two consecutive iterations.
In this work, $N=30$ CTs were selected for the optimization process. The order of complexity for exhaustive search requires $\mathcal{O}(2^N)$, where as the ADSE provides optimal CTPs with a complexity of $\mathcal{O}(N \cdot I)$, depending on the number of iterations $I$ \cite{dse}.

\section{Extension of the Advanced Design Space Exploration}
\label{sec:extension}

Prior to this work, a CT is either enabled or disabled for the entire bitstream. In contrast, the proposed approach offers a more granular switching of CT usage on frame-level.
%Thereby, it is not only possible to find new rate-energy-efficiency trade-offs, but also the overall Pareto curve can be enhanced. To enable such a more granular CT usage, the VVenC encoder is modified as described in Subsection \ref{subsec:VVCEncoder}.
For this purpose, the VVenC encoder is modified, as described in Subsection \ref{subsec:VVCEncoder}.
 Subsequently, the ADSE optimzation algorithm from Section \ref{sec:ADSE} is extended to search the enlarged space of CTPs. The extension is referred to as continuous advanced design space exploration (CADSE) and described in Subsection \ref{subsec:CADSE}.
Finally, new optimization criteria are introduced in Subsection \ref{subsec:Costfkt}.

\subsection{Modification of the VVC Encoder}
\label{subsec:VVCEncoder}

Instead of a binary on/off state, each coding tool is assigned to a dynamic tool rate $r(\nu)$ which is defined to be in range,
\begin{equation}
	 r(\nu)= m \cdot 0.125 \, \text{ for } m \in \mathbb{N}_0 < 9.
\end{equation}
This tool rate dynamically defines for which frames the CT $\nu$ shall be enabled or disabled respectively.
In fact, this more granular definition of CT usage exploits the repetitive group of pictures (GOP) structure in VVC.
For the \textit{randomaccess} configuration, a GOP consists of 32 frames that spread over 6 temporal layers \cite{JVET}. The GOP structure defines a temporal hierarchy for inter-prediction. Frames on higher temporal layers depend on frames on the lower temporal layers. Consequently, frames on the lower temporal layers should generally exhibit a good visual quality.
With this relation in mind, the dynamic tool rate is also defined for one GOP and will then be repeated in this pattern for the entire bitstream.
%Maximum granularity can be achieved by choosing a step size of $ \frac{1}{32}$ for the dynamic rate.
Choosing a step size of 0.125 controls $ 0.125 \cdot 32 = 4$ frames at a time.
In future experiments the granularity can be maximized by choosing a step size of $ \frac{1}{32}$. 
A tool rate of $r(\nu) = 1$ means that the CT of interest is enabled for all frames in one GOP.
%Decreasing the rate by 0.125, disables the CT usage in four frames on the highest temporal layer.  
Each tool rate decrease by 0.125 reduces the number of frames in which the CT is enabled by four.
The tool usage is first disabled in the frames on the higher temporal layers.
A tool rate equal to $r(\nu) = 0$ relates to a CT which is disabled in the entire GOP and therefore in the entire bitstream.
The modified VVenC encoder is adjusted, to accept such coding tool rates in the config-files. 
%In order to adapt the usage of a CT on frame-level, the CT state at the beginning of the encoding process of a frame is accessed.
%Depending on the position of a frame in the GOP, and the configured coding tool rate $r(\nu)$, the coding tool state is modified and redefined for the current frame.
For each frame the position within the GOP is identified using the picture oder count (POC) and depending on the configured coding tool rate $r(\nu)$, the coding tool usage is enabled or disabled accordingly for each frame.

%Todo: kÜrzen???
%Implementation-wise it should be noted that such a modification of the tool states on frame level is initially not intended by the encoder. Therefore, the original sequence parameter set (SPS) that keeps track of the CT states is defined as a read-only struct. This is bypassed by creating an exact copy of the SPS struct which is not defined as a \texttt{const}, thus values may be changed at any point. All variable calls in the code, where the CT state is accessed are replaced with a reference to the new SPS structure.
Implementation-wise, special attention has to be paid regarding the occuring tool combinations to guarantee decodability for each frame and to remain compliant to the standard. 
Since each CT can operate at its own rate, invalid combinations have to be intercepted for each frame. For example, it has to be validated that Block-DPCM is only enabled in combination with an activated transform skip mode.  

 %\vspace{-0.1cm}
\subsection{Continuous Advanced Design Space Exploration}
\label{subsec:CADSE}
Each of the nine coding tool rates $ r(\nu)$ corresponds to a different fraction of frames for which a CT is enabled. Consequently, the number of possible configurations rises significantly and conducting a full search would require an order of complexity of $\mathcal{O}(9^N)$. However, it is possible to extend the ADSE to search and navigate through the enlarged search space without increasing the complexity of the algorithm.
The extended optimization, namely CADSE, basically operates in the same iterative fashion as ADSE (c.f., Section \ref{sec:ADSE}). With the major difference that no longer a binary flip of the tool usage is performed to test the impact of CTs isolated, but instead a change of the dynamic coding tool rate $r$ is derived.

As before in ADSE, a binary CTP is used as baseline and initial reference. In the first iteration, each tested CTP $\boldsymbol{u}_{1,\nu}$ is a copy of the reference CTP except for the entry at $\boldsymbol{u}_{1,\nu}(\nu) = 0.5 $, which basically tests the CT $\nu$ isolated at a rate of $r(\nu)=0.5$. 
These CTPs are used to encode the test sequences, and during decoding the energy consumption as well as the resulting quality are measured. Following the selection process from \eqref{eq:select}, CTPs that lead to a joint efficiency improvement are identified.
The CT usage rates which proved to be beneficial are combined into the new reference CTP for the next iteration. In the second iteration $(i =2)$, a second set $\mathbf{U}_i$ of CTPs for testing has to be derived from the new reference. To this end, copies of the reference $\boldsymbol{u}_{2,0}$ are taken, and in each CTP for testing, the entry of the $\nu$-th tool is modified by,
\vspace{-0.3cm}
\begin{align}
 	\boldsymbol{u}_{i, \nu} &= \boldsymbol{u}_{i,0} \\
	 \boldsymbol{u}_{i, \nu}(\nu) &= \boldsymbol{u}_{i,0}(\nu) + \Delta r(\nu),
	 \label{eq:add}
\end{align}
where $\Delta r(\nu)$ denotes a tool rate change that is determined in each iteration for each CT individually.

To compute this rate change, multiple parameters are considered.
First, the efficiency difference of the previous reference CTP $\boldsymbol{u}_{i-1,0}$ and the previously tested CTPs $\boldsymbol{u}_{i-1,\nu}$ are evaluated in $\boldsymbol{e}_i$ which is in turn used to derive the amount of rate change $|\Delta r(\nu)|$ by,
 \begin{align}
 	\boldsymbol{e}_i(\nu) &= f(\boldsymbol{u}_{i-1,0}) - f(\boldsymbol{u}_{i-1,\nu})  \\
 	|\Delta r(\nu)| &= \min \left( \frac{\lceil |\mathbf{e}_i(\nu)| \cdot 0.4 \rceil}{8} , 0.5 \right ).
 	\label{eq:ratechnge}
 \end{align}
 Provided that $\boldsymbol{e}_i(\nu) \neq 0\%$ holds, four rate changes of different strength occur which are all multiples of 0.125, according to \eqref{eq:ratechnge}. Considering the range of the efficiency differences, a maximum rate change of $|\Delta r(\nu)| = 0.5\%$ is exemplarily obtained by $\boldsymbol{e}_i(\nu)= 10\%$.
  Secondly, the direction of the previous CT test is evaluated in $\boldsymbol{t}_i$, and finally, the development of the CT usage in the two consecutive reference configurations is compared and denoted in $\boldsymbol{d}_i$, following,
  \begin{align}
	\boldsymbol{t}_i(\nu) &= \text{sign}(\boldsymbol{u}_{i-1,0}(\nu) - \boldsymbol{u}_{i-1,\nu}(\nu)) \\
	\boldsymbol{d}_i(\nu) &= \text{sign}(\boldsymbol{u}_{i-1,0}(\nu) - \boldsymbol{u}_{i,0}(\nu)).
	\label{eq:direction}
\end{align}
Combining all of the acquired information leads to two cases for the tool rate change,
\begin{align}
	\Delta r(\nu) =
\begin{cases}
|\Delta r(\nu)|\cdot \boldsymbol{t}_i(\nu) \cdot -1 \, \, &\text { if } \, \mathbf{d}_i(\nu) = 0,\\
|\Delta r(\nu)|\cdot \boldsymbol{t}_i(\nu) \, \, &\text { if } \, \boldsymbol{d}_i(\nu) \neq 0.\\
\end{cases}
\end{align}
The first case ($\boldsymbol{d}_i(\nu) = 0$) covers the situation that a tested CT usage is less efficient than the reference, the tool usage is therefore identical in both consecutive reference configurations. 
Consequently, the tool performance shall be tested, when applying a rate change in the opposite direction than the previous test.
In the second case ($\boldsymbol{d}_i(\nu) \neq 0$), the tested CT usage is identified to be efficient. As a result, the tool usage is increased in the same direction as in the former iteration.

Following \eqref{eq:add}, the respective rate change is used to derive the next set of CTPs for testing. This procedure is repeated in each iteration and enables the search of efficient CTPs within the enlarged, and more continuous search space.
Again, the algorithm terminates if identical reference configurations are identified in consecutive iterations.
%\vspace{-0.1cm}
\subsection{Enhancement of the Cost Function}
\label{subsec:Costfkt}

When jointly optimizing decoding energy and compression, so far the sum from \eqref{eq:classicCostfunction} has been used. This optimization criterion weights the increase in rate and the decrease in decoding energy equally. However, in some applications the acceptable amount of additional bitrate may be limited, and the maximum decoding energy reduction in the region around this limit $l$ is of interest.
Therefore, optimization criteria are proposed and investigated.
The linear criterion is designed following
\begin{equation}
		f_1(\boldsymbol{u}) = \max(\text{BDR}_{\text{VMAF}}-l, 0\%)\cdot w + \text{BDDE}_{\text{VMAF}},
		\label{eq:linear}
\end{equation}
which neglects the cost of BDR up-to the limit $l$ and then penalizes higher bit rate requirements linearly with a factor of $w$, which is in this work evaluated at $w=3$.
In contrast, the linear-cubic criterion distributes the weight following
\begin{equation}
\begin{split}
		f_2(\boldsymbol{u}) &= \min(\text{BDR}_{\text{VMAF}}-(l-b),0\%) \\
		&+ \max(\text{BDR}_{\text{VMAF}}-(l+b),0\%)^3\\
		&+ \text{BDDE}_{\text{VMAF}}.
		\label{eq:linCubic}
\end{split}
\end{equation}
Consequently, CTPs which lead to trade-offs with few additional bits are favoured, the region $[l-b, l+b]$ around the defined bite rate limit $l$ is equally weighted as BDDE, and higher rate requirements are penalized cubically. 
In this work, the linear and the linear-cubic cost function are designed to explore the region around the limit $l=5\%$ BDR and for $f_2(\boldsymbol{u})$ the equal weighting of BDR and BDDE is restricted by the bounds $b = 0.5\%$ .

\section{Setup and metrics}
\label{sec:setup}

For the evaluation with ADSE, the optimized software encoder implementation VVenC in version 1.7.0 is used \cite{vvencRepo}. Correspondingly, VVdeC is used for the decoding procedure in the hitherto latest version 1.6.1 \cite{vvdecRepo}.
The granular optimization approach, which is presented in this paper, uses the same decoder in order to achieve a fair comparison regarding the decoding energy.
In this work, the visual quality of the decoded bitstreams is assessed with VMAF, since it matches the perceived visual quality of a human better than PSNR \cite{vmafRepo}, \cite{VMAF}.
The decoding energy measurements were conducted on a desktop PC that runs CentOs Stream 8 as an operating system. The build in processor is an Intel i5-4670 CPU that has a x86 architecture and consists of four cores which operate at a base frequency of 3.4GHz. Energy measurements on the PC are conducted with the Running Average Power Limit (RAPL), which directly measures the power demand of the CPU \cite{RAPL}. 
To ensure statistical correctness, a measurement is repeated multiple times and  the results are verified with a confidence interval test, as described in \cite{dse}.
A selection of five video sequences is taken from the VTM common test condition \cite{TestSet}. All the sequences are HD videos, belonging to class B. To determine the efficiency of a CTP, the configuration under test is used to encode the first 128 frames of the five sequences with four different quantization parameters (QP) $\in [22,27,32,37]$.
To evaluate the energy and compression efficiency, the Bj\o ntegaard-Delta (BD) metric based on the description from \cite{dse} is used. The BD metric is a comparative measure that indicates increase or decrease in percent compared to a reference \cite{BD}.
Throughout this work the \textit{slower} preset of VVenC is the reference for all BD calculations. 
$\text{BDR}_{\text{VMAF}}$ indicates the bit rate increase or decrease for the same visual quality and likewise $\text{BDDE}_{\text{VMAF}}$ provides insight about the reduction of the decoding energy for an equal VMAF score.

In the course of this work, the same 30 CTs as in \cite{adse} are analysed. Granular switching is implemented and enabled for all of them with the exception of IBC, MTS and SbTMVP. The excluded CTs have to be further investigated with respect to their relation and influence on other CTs. 
%\vspace{-0.1cm}
\section{Results and Evaluation}
\label{sec:results}
\subsection{Derivation of CTPs with ADSE}
\label{subsec:ResultsADSE}

\begin{figure}[hb!]
%\vspace{0.2cm}
\centering
\input{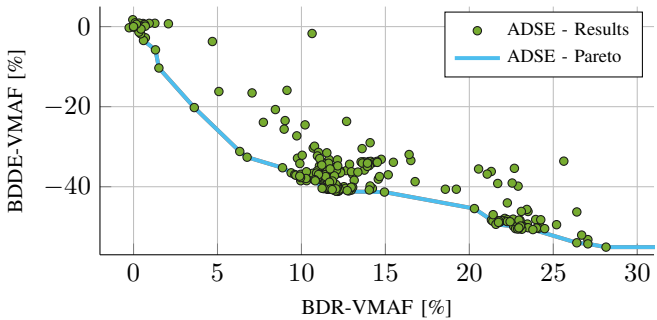}
\vspace{-0.4cm}
\caption{Overview of the efficiency trade-offs resulting from jointly optimizing decoding energy and bit rate with ADSE. In cases where only up to 10\% additional bit rate is acceptable, very few CTPs offer a trade-off. Also the region between the dominant clusters is sparsely populated.}
\label{fig:ADSE}
\end{figure}

The ADSE algorithm from the literature searches for optimal CTPs that define the CT usage for the entire bitstream. The achievable efficiency trade-offs are shown in Fig. \ref{fig:ADSE}, where each marker corresponds to the efficiency properties of a tested CTP. The vertical axis indicates the amount of energy reduction as $\text{BDDE}_{\text{VMAF}}$ and the horizontal axis shows the bit rate increase in $\text{BDR}_{\text{VMAF}}$.
The starting point of ADSE is the VVenC \textit{slower} configuration, located in the top-left origin of the diagram.
Over the course of nine iterations, two dominant clusters become apparent.
The first dominant cluster is at around 10\%-15\% $\text{BDR}_{\text{VMAF}}$, where the CTPs provide a reduction of around 40\% decoding energy compared to the VVenC \textit{slower} preset. 
The second cluster is found in the region of 23\% $\text{BDR}_{\text{VMAF}}$ and -50\% $\text{BDDE}_{\text{VMAF}}$.
A Pareto front is defined by all CTP points that provide maximum compression for a given decoding energy reduction.

\captionsetup[figure]{font=footnotesize}
\captionsetup{belowskip=-15pt}
\vspace{0.3cm}

The Pareto front in Fig. \ref{fig:ADSE} thereby units all of the tested ADSE CTPs for which the decoding energy cannot be further reduced without increasing the bit rate. Yet, it can be observed that there remain gaps where the Pareto curve is interpolated over large sections.
In these sections, none of the tested binary CTPs can provide an efficiency trade-off.

\begin{center}
\captionsetup[figure]{font=footnotesize}
\captionsetup{belowskip=-15pt}
\begin{figure*}[htbp]
\vspace{-0.5 cm}
\centering
\input{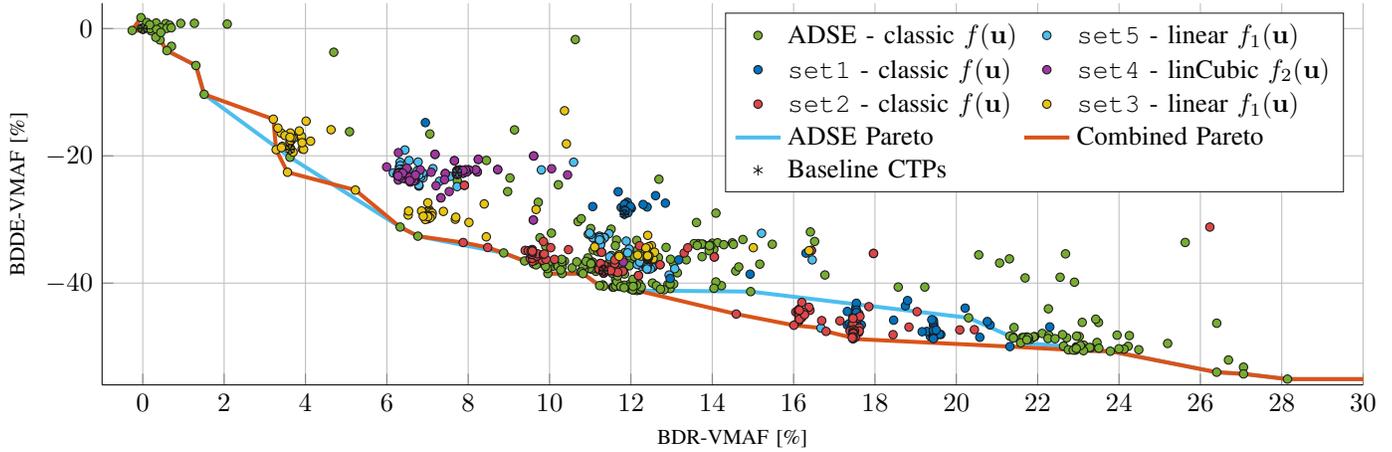}
\vspace{-0.5cm}
\caption{Extension of the design space by CTPs derived with CADSE. Each marker represents a CTP. The baseline CTP of each set is highlighted with an asterisk. - Zoom in for details.}
\label{fig:AllResults}
\end{figure*}
\end{center}

\vspace{-0.85cm}
\subsection{Derivation of CTPs with CADSE compared to ADSE}
\label{subsec:ResultsCADSE}
To demonstrate that CADSE is able to fine-tune and improve the efficiency of CTPs, four binary baseline CTPs that already decrease the decoding energy demand, compared to the \textit{randomaccess slower} profile, are selected.
Starting at the four respective baseline CTPs, five sets were tested with the different cost functions from Section \ref{subsec:Costfkt}.
An overview of the sets and the properties of their baseline CTPs is given in Table \ref{tab:baseline}.
Additionally, Table \ref{tab:baseline} lists for each set a CTP with dynamic tool usage that was found to improve the respective baseline CTP.
The efficiency properties of the baseline and the improved CTPs are listed as tuples with $\text{BDR}_{\text{VMAF}}$ on top and $\text{BDDE}_{\text{VMAF}}$ below.

In Fig. \ref{fig:AllResults} the iterative development of all sets is illustrated. 
%Each marker point corresponds to a tested CTP. 
%The CTPs investigated by CADSE are derived as described in Section \ref{subsec:CADSE} and have continuous tool rates.
Due to the granular adjustment of CTs on frame-level, novel efficiency trade-offs are found in the design space.
By introducing the dynamic tool rates, CADSE is visibly able to improve the efficiency of the respective baseline CTPs. On top, these improvements are already determined after the first few iterations of each set.
It can be seen that the new, combined Pareto front (red) unites more CTPs with actual efficiency trade-offs. Consequently, the interpolated sections of the curve become smaller and the continuity is significantly improved.

 In this work, the cost functions  explore the region around 5\% BDR, and indeed
 $f_1(\boldsymbol{u})$ and $f_2(\boldsymbol{u})$ shift the focus of the optimization in the direction of this specified bit rate.
 The difference  between the linear-cubic and the linear cost function is emphasized by the comparison of \texttt{set4} and \texttt{set5}, which both fine-tune the same baseline CTP.
  It can be seen that the linear criterion $f_1(\boldsymbol{u})$ used in \texttt{set5} (light-blue) does not only yield CTPs in the low bit-rate range.
This can be explained with the smaller increase in cost for higher bit rates compared to the linear-cubic cost function $f_2(\boldsymbol{u})$ (purple).
Together with the fine-tuning of the baseline CTP of the linear \texttt{set3}, it can be seen that CADSE combined with the cost-functions is able to fill the lower bit-rate range with CTPs.

     \begin{table}[!b]  
  \captionsetup{font=footnotesize}      
%\vspace{+0.3cm}
\centering
\renewcommand{\arraystretch}{1.2}
	\begin{adjustbox}{max width=\textwidth}
       %\subfloat[Baseline CTPs for CADSE]{
       \begin{tabular}{l || c : c : c : c :c } 
         %& \multicolumn{2}{c|}{EE} &  \multicolumn{2}{c}{EBE}  \\
         & \texttt{set1} & \texttt{set2} & \texttt{set3} & \texttt{set4} & \texttt{set5}    \\
         \hline 
         %Cost function & classic & classic & linear & linear & linear-cubic \\
 		 \centering Cost-  & $f(\boldsymbol{u})$ &$f(\boldsymbol{u})$ & $f_1(\boldsymbol{u})$ & $f_2(\boldsymbol{u})$ & $f_1(\boldsymbol{u})$ \\
 		 \centering function &  \footnotesize{\textcolor{gray!90}{classic}}& \footnotesize{\textcolor{gray!90}{classic}} & \footnotesize{\textcolor{gray!90}{linear}} & \footnotesize{\textcolor{gray!90}{linear-cubic}} & \footnotesize{\textcolor{gray!90}{linear}} \\
 		 \hline
 		 Baseline & 11.84\,\%  & 11.33\,\%  & 3.62\,\%   & 7.73\,\%  &  7.73\,\%  \\
 		 
 		&   -29.16\,\% &  -38.03\,\% & -18.99\,\% & -22.51\,\%  & -22.51\,\% \\
 		 \hline
 		 Improved & 21.32\,\% & 17.46\,\%  & 3.56\,\%  & 6.28\,\%   & 7.32\,\%   \\
 		   &  -49.98\,\%  & -48.77\,\%  & -22.58\,\%  & -23.81\,\% & -26.60\,\%  \\
 		
       \end{tabular}
          %    }
       \end{adjustbox} 
 
\caption{Overview of the tested sets with the cost-functions. The efficency properties of the baseline CTPs [$\text{BDR}_{\text{VMAF}}$ on top of $\text{BDDE}_{\text{VMAF}}$] are given for each set. The table also lists the improved efficiency properties of an exemplary CTP which was derived during the fine-tuning process of each set.}
       \label{tab:baseline}
     \end{table}     

 Further, CADSE offers new pareto-optimal efficiency trade-offs with the classical optimization criterion $f(\boldsymbol{u})$, which has also been used for ADSE.
 Both, \texttt{set1} (blue) and \texttt{set2} (red) optimize with $f(\boldsymbol{u})$, but fine-tune slightly different baseline CTPs. Yet, both sets derive CTPs with efficiency properties in the empty region between the two dominant ADSE clusters (12\%-22\%). In fact, in this region the Pareto front is improved by around 5\% BDDE.
The configuration that yields the best overall joint rate-energy efficiency (BDR+BDDE) is obtained by \texttt{set2}.
For the tested JVET Class B video sequences, this CTP decreases the decoding energy by 48.77\% when spending 17.46\% additional bit rate.
With ADSE, a comparable energy reduction of 48.81\%, was found at the cost of a bit-rate increase by 21.83\%.
Comparing these efficiency properties, it can be seen that CADSE offers savings by more than 4\% in $\text{BDR}_{\text{VMAF}}$ compared to ADSE.

\vspace{-0.05cm}
\section{Conclusion and Outlook on Future Work}
\label{sec:conclusion}
In this paper we integrate control over the coding tool usage on frame-level in the state-of-the-art VVenC coder.
%As a result the number of possible CTPs increases several times over.
We present an algorithm that derives CTPs that optimze decoding energy and rate in VVC jointly by operating on frame-level.
With the proposed approach, binary CTPs can be fine-tuned within few additional iterations and their efficiency properties can be enhanced. These new trade-offs extend the design space and the Pareto front becomes more continuous.

In addition, the proposed methodology provides a significant improvement of the Pareto front. At 17.46\% $\text{BDR}_{\text{VMAF}}$, a reduction of 48.77\% of the decoding energy demand can be achieved. This reduction of energy requires 4.3\% less additional bit rate than with the state-of-the-art  approach. 
Thereby, CADSE contributes to achieving more frugality for video streaming by extending the set of pareto-efficient CTPs. With such a set at hand, it is possible to choose an efficient CTP from the set, which matches any user-specified efficiency properties.
%This contributes to achieving more frugality in video streaming by deriving a large set of optimal CTPs, where each CTP satisfies a different efficiency trade-off between reduced decoding energy demand and increased bit rate. 
In the future, we plan to implement granular switching for all of the 30 CTs from \cite{adse}. Moreover, we strive to repeat the presented methodology with a combination of video sequences from all JVET classes, instead of only class B sequences. We expect this to provide universally valid CTPs that are robust against changes in content. Finally, the cost functions proved to bear potential and further adjustments will be investigated to narrow down the region of interest in the design space even more.

%Enable granular switching on frame-level for all CTs in VVC. 

%Further testing of costfunctions.

%TODO: Outlook - test dont treat all gops identical

\bibliographystyle{IEEEtran}
\bibliography{referencesEuSipCo}

\end{document}